\documentclass[aps,preprint,showpacs,superscriptaddress]{revtex4}
\usepackage{rotating}
\usepackage{graphicx}
\usepackage{setspace}
\usepackage{epsfig}
\usepackage{dcolumn}
\usepackage{bm}
\newcommand{\be}{\begin{equation}}
\newcommand{\ee}{\end{equation}}
\newcommand{\bear}{\begin{eqnarray}}
\newcommand{\eear}{\end{eqnarray}}
\newcommand{\lomega}{\ensuremath{\lambda_\omega}}
\begin{document}
\title{\Large \bf Pion correlations in Nuclear Matter}
\author{P.K. Panda,}
\address{Centro de F\'isica Computacional, Departamento de F\'isica,
Universidade de Coimbra, P-3004 - 516 Coimbra, Portugal}
\author{S. Sarangi,}
\address{ICFAI Institute of Science \& Technology, 
Bhubaneswar-751 010, India}
\author{J. da Provid\^encia}
\address{Centro de F\'isica Computacional, Departamento de F\'isica,
Universidade de Coimbra, P-3004 - 516 Coimbra, Portugal}
\begin{abstract}
The saturation properties of the nuclear matter taking pion correlations 
into account is 
studied. We construct a Bogoliubov transformations for the pion pair
operators and calculate the energy associated with the pion pairs. 
The pion dispersion relation is investigated. We next study the correlation 
energy due to one pion exchange in nuclear matter and neutron matter at random
phase approximation using the generator coordinate method. The techniques 
of the charged pion correlations are discussed in the neutron matter 
calculations. We observe that there is no sign of the pion condensation
in this model.
\end{abstract}

\maketitle

\section {Introduction}
The understanding of the nuclear force at a microscopic level is an important 
problem since it shall be the basis for the nuclear matter and finite nuclei 
calculations. The interactions of nucleons which may arise as a residual 
interaction due to their substructure of quarks and gluons is technically not 
solvable. The alternative approach is to tackle the problem through meson 
interactions.

In recent years relativistic mean field theory (RMF)\cite{Walecka74,Serot86} 
has been quite successful in describing nuclear matter and finite nuclei 
properties.  The NN dynamics arises in this model from the exchange of a 
Lorentz scalar isoscalar meson, $\sigma$, which provides the mid range 
attraction and an  isoscalar vector meson, $\omega$, which provides 
the repulsion. Also, in this 
model the inclusion of the $\rho$- meson takes care of the neutron-proton 
asymmetry. With a small number of parameters, the RMF model reproduces the 
nuclear matter saturation and describes the bulk and the single particle 
properties for the finite nuclei reasonably well.

Despite the success of the RMF model, several open questions still remain 
unanswered. Firstly, the meson fields are classical and secondly the attractive 
part of the nuclear force is mediated through the hypothetical $\sigma$ meson 
which could be an effect of multi-pion exchanges 
\cite{mishra90,mishra92,panda92,panda96,sarangi08}. The $\sigma$ can not be 
interpreted as the representation of a physical particle, since such a 
particle or resonant state remains still to be confirmed in the experiment. 
Thus on aesthetic as well as phenomenological ground, alternative approaches 
will add to our understandings.

The original Walecka model at the Hartree approximation does not contain 
a dynamical description of the pion
fields. However the importance of the pions in NN dynamics can not be ignored. 
Realizing the essential role of pion in the description of the 
nuclear medium an alternative approach for the nuclear matter \cite{mishra92},
deuteron \cite{panda92} and for $^4$He \cite{panda96} has been developed. 
In this method, it has been studied the description of nuclear matter 
using pion pairs through a squeezed coherent state type 
of construction \cite{mishra92,panda92,panda96,sarangi08}. This simulates 
the effects of the $\sigma$-meson and is a 
very natural quantum mechanical formalism for the classical fields.

The generator coordinate method (GCM) is a technique of great physical appeal 
which has been developed \cite{hill53} to describe collective oscillations
in nuclei. Besides being extensively used in nuclear structure physics,
it often finds application in various other branches of physics 
\cite{johansson78}. In this work we investigate the pion condensation problem
in nuclear matter and neutron matter using the generator coordinate method.
In a simplified model, the problem had been studied in Ref. \cite{providencia77}
where a coherent state description for the pions was used.
Using the gaussian overlap and harmonic approximation,
the Hamiltonian may be diagonalized by a randam phase approximation (RPA) like
canonical transformation \cite{chatto81}. In this methodology, one can go 
beyond the coherent state description. Again the calculation can be 
carried out exactly, without further approximation.
The present analysis is an extension of 
the mean field approach of Walecka where classical fields are replaced by 
quantum coherent states for the pion pairs. This report has also an 
advantage that the one pion exchange correlation contributions are  
considered at RPA level using similar Bogoliubov transformations. In the 
present model, we have observed no sign of the pion condensations in the 
RPA modes.

The outline of the paper follows. In section 2, we derive a pion nucleon 
Hamiltonian in a non-relativistic limit. We then construct a Bogoliubov 
transformations for the pion pair operators and calculate the energy 
associated with the pion pairs. In section 3, we calculate the correlation 
energies due to one pion exchange in nuclear matter and neutron matter at 
random phase approximation (RPA).  Section 4 consists of the discussions of 
the saturation properties of nuclear matter, the pion dispersion relation in 
the medium and a concluding remarks. 
\section{Formalism}
\label{Formalism}
\subsection{Non-relativistic Hamiltonian}
The Lagrangian for the pion nucleon system is taken as
\begin{equation}
{\cal L}=\bar{\psi}\left (i\gamma^{\mu} \partial_{\mu}-M+G\gamma_5\phi
\right )\psi- \frac{1}{2}\left(\partial_{\mu}\varphi_i\partial^{\mu}\varphi_i-
m^2\varphi_i\varphi_i\right ),\label{e1}
\end{equation}
where $\psi=\left (
\begin{array}{c}
\psi_I\\
\psi_{II}\\
\end{array}
\right )$ is the doublet of the nucleon field
with mass $M$, $\varphi_i$'s are pion fields and $\phi = \tau_i\varphi_i$ 
represents the off-mass shell isospin triplet pion field with mass $m$. 
$G$ is the pion-nucleon coupling constant. Repeated indices indicate summation.

The representations of $\gamma$ matrices are
\[\vec\gamma =
\left (
\begin{array}{cr}
0 & \vec\sigma\\
-\vec\sigma & 0\\
\end{array}
\right ),\quad\quad
\gamma_0 =\left (
\begin{array}{cr}
1 & 0\\
0 & -1\\
\end{array}\right ),\quad\quad
\gamma_5 =\left (
\begin{array}{cr}
0 & -i\\
-i & 0\\
\end{array}\right )\ .
\]
From the above Lagrangian, the equation of motions are
\begin{equation}
(E-M)\psi_I-(\vec\sigma\cdot \vec p+iG\phi)\psi_{II}=0\ ,
\end{equation}
\begin{equation}
(E+M)\psi_{II}-(\vec\sigma\cdot\vec p-iG\phi)\psi_I=0\ ,
\end{equation}
where, $E=i(\partial/\partial t)$ and $\vec p=i(\partial/\partial \vec x)$. 
Eliminating the small component $\psi_{II}$ from equation (2) and (3) we have
\begin{equation}
\left[(E^2- M^2)-(E+M)(\vec\sigma\cdot p+iG\phi)(E+M)^{-1}
(\vec\sigma\cdot\vec p-iG\phi)\right]\psi_I=0\ .
\end{equation}
Equation (4) can be rewritten as
\begin{equation}
\left[E^2- M^2-p^2+iG[(\vec\sigma\cdot\vec p),\phi]-G^2\phi\cdot \phi\right]
\psi_I=0\ .
\end{equation}
From equation (5), we can immediately identify the effective Hamiltonian for 
the nucleons as
\begin{eqnarray}
{\cal H}_N&=&\psi_I^\dagger(\vec x)\left[p^2+ M^2-iG
((\vec\sigma\cdot\vec p)\phi)+G^2\phi^2\right]^{1/2}\psi_I(\vec x)\nonumber\\
&\simeq&\psi_I^\dagger(\vec x)\left[\epsilon_x-\frac{iG}{2\epsilon_x}
((\vec\sigma\cdot\vec p)\phi)+\frac{G^2}{2\epsilon_x}\phi^2\right]
\psi_I(\vec x)\nonumber\\
&=&{\cal H}^0_N(\mathbf x)+{\cal H}_{int}(\mathbf x)\ ,
\end{eqnarray}
where the single particle nucleon energy operator $\epsilon_x$
is given by $\epsilon_x=(M^2-\vec\nabla_x^2)^{1/2}$. In the 
non-relativistic assumption, we have to replace  $\epsilon_x$ by $M$, 
when in a denominator and by $M+\frac{p^2}{2M}$ when not in a denominator.
Now the effective Hamiltonian becomes
\begin{equation}
{\cal H}(\mathbf x)={\cal H}^0_N(\mathbf x)+{\cal H}_{int}(\mathbf x)+
{\cal H}_M({\mathbf x}),\label{e2}
\end{equation}
where the free nucleon part ${\cal H}^0_N (\mathbf x)$ is given by
\begin{equation}
{\cal H}^0_N(\mathbf x)=\psi^\dagger(\mathbf x)~\left(M+
\frac{\nabla_x^2}{2M}\right)~\psi(\mathbf x)\ ,\label{e3}
\end{equation}
the $\pi N$ interaction Hamiltonian is provided by
\begin{equation}
{\cal H}_{int}(\mathbf x)=\psi^\dagger(\mathbf x) \left[
-{iG\over 2 M }((\mbox{\boldmath$\sigma$}\cdot \mathbf p) ~\phi) +
{G^2\over 2 M }\phi^2\right]\psi(\mathbf x)\ ,\label{e5}
\end{equation}
and the free meson part ${\cal H}_M(\mathbf x)$ is defined as
\begin{equation}
{\cal H}_M(\mathbf x)={1\over 2}\left[{\dot \varphi}_i^2
+(\mbox{\boldmath $\nabla$}\varphi_i)\cdot(\mbox{\boldmath$\nabla$}\varphi_i)
+m^2\varphi_i^2\right]\ .\label{e4}
\end{equation}
We expand the pion field operator $\varphi_i(\mathbf x)$ in terms
of the creation and annihilation operators of off-mass shell pions
satisfying equal time algebra as
\begin{equation}
\varphi_i(\mathbf x)={1\over \sqrt{2 \omega_x}}
(a_i(\mathbf x)^\dagger +a_i(\mathbf x)),~~~~~~~~~~
\dot\varphi_i(\mathbf x)=i{\sqrt{\omega_x\over 2}}
(a_i(\mathbf x)^\dagger -a_i(\mathbf x))\ ,\label{e6}
\end{equation}
with energy $\omega_x = (m^2-\mbox{\boldmath $\nabla$}_x^2)^{1/2}$.

\subsection{Correlation energy associated with two pions and Bogoliubov Transformation}
The quadratic terms in the pion field in eq.~(\ref{e5}) provide a
isoscalar scalar interaction of nucleons and thus would simulate the effects
of $\sigma$-mesons of the Walecka model.

A pion-pair creation operator given as
\begin{equation}
B^{\dag} =\frac{1}{2}\sum_{\mathbf k} f_{\mathbf k}~a_{\mathbf k i}^{\dag}~a_{
-\mathbf k i}^{\dag},
\label{e7}
\end{equation}
is then constructed with the creation and annihilation operators in momentum space and
the ansatz function ${f}(\mathbf k)$. We then define the unitary transformation $U$ as
\begin{equation}
U=e^{(B^{\dag}-B)}\label{e8}
\end{equation}
and note that $U$, operating on vacuum, creates an arbitrarily large number of
scalar isospin singlet pairs of pions corresponding to squeezed coherent states.
We will show that this is the appropriate transformation to diagonalize the
pion part of the Hamiltonian.
The ``pion dressing'' of nuclear matter is then introduced through the state
\begin{equation}
|\Psi\rangle=U|0\rangle=e^{(B^\dagger-B)}|0\rangle.\label{e9}
\end{equation}
We obtain
\begin{equation}
\tilde a_{\mathbf k i}=U^\dagger~a_{\mathbf k i}U=(\cosh
f_{\mathbf k})~a_{\mathbf ki}+
(\sinh f_{\mathbf k})~a_{-\mathbf k i}^\dagger,
\label{e11}
\end{equation}
which is a Bogoliubov transformation. Here $U$ is a unitary and hermitian 
operator. The psedo-pions
$\tilde a_{\mathbf k i}$ are the results of the unitary transformation. 
It can also be easily checked that the operator $\tilde a_{\mathbf k i}$ 
satisfies the standard bosonic commutation relations:
\begin{equation}
[\tilde a_{\mathbf{k} i},~\tilde a^{\dag}_{\mathbf{k'}j}]=\delta_{ij}
\delta_{\mathbf{k},\mathbf{k'}},~~~~
[\tilde a^{\dag}_{\mathbf{k} i},~\tilde a^{\dag}_{\mathbf{k'} j}]
=[\tilde a_{\mathbf{k}i},~\tilde a_{\mathbf{k'}j}]=0\ .
\label{e12}
\end{equation}
and also
\begin{equation}
\tilde a_{\mathbf k i}|\Psi\rangle=0
\end{equation}
The reverse transformation:
\begin{equation}
a_{\mathbf k i}=(\cosh\ f_{\mathbf k})~\tilde a_{\mathbf k i}-
(\sinh\ f_{\mathbf k})~\tilde a_{-\mathbf k i}^\dagger
\equiv x_{\mathbf k}~\tilde a_{\mathbf k i}-
y_{\mathbf k}~\tilde a_{-\mathbf k i}^\dagger\ .
\label{e13}
\end{equation}
In momentum space the effective Hamiltonian (\ref{e2}) may be re-written as
\begin{eqnarray}
H&\approx&\sum_{\mathbf p,\alpha\eta}\varepsilon_p~ c^\dagger_{\mathbf p,
\alpha\eta}c_{\mathbf p,\alpha\eta}
+\sum_{\mathbf q,j}\omega_{\mathbf q} a^\dagger_{\mathbf q,j}a_{\mathbf q,j}\nonumber\\
&-&\sum_{\mathbf p \mathbf q,j\alpha\alpha'\eta\eta'}\frac{G}{2M\sqrt{\omega_
{\mathbf q}V}}~c^\dagger_{\mathbf p+\mathbf q,\alpha\eta}c_{\mathbf p,\alpha'
\eta'}(i\sigma.\mathbf q)_{\alpha\alpha'}\tau_j(a_{\mathbf q,j}+
a^\dagger_{-\mathbf q,j})\nonumber\\
&+&\sum_{\mathbf p\mathbf q,j\alpha\eta}\frac{G^2}{2M\omega_{
\mathbf q}V}~c^\dagger_{\mathbf p,\alpha\eta}c_{\mathbf p,\alpha\eta}\left(
a^\dagger_{\mathbf q,j}a^\dagger_{-\mathbf q,j}
+a_{\mathbf q,j}a_{-\mathbf q,j}+2a^\dagger_{\mathbf q,j}a_{\mathbf q,j}\right).
\label{eqHam}
\end{eqnarray}
Here $\mathbf p$, $\alpha$ and $\eta$ are respectively, the momentum, spin and 
iso-spin quantum numbers of the nucleon and $\mathbf q$, $j$ are the momentum 
and isospin labels of the pion. $p=|\mathbf p|$, $q=|\mathbf q|$.
$c^\dagger_{\mathbf p,\alpha\eta}$ is the creation operator for nucleon with 
momentum $\mathbf p$, spin $\alpha$ and isospin $\eta$.
The contribution of the quadratic term in the pion field coming from the 
above Hamiltonian as
\begin{eqnarray}
H_{2\pi}&=&\sum_{\mathbf q,j}\omega_{\mathbf q} a^\dagger_{\mathbf q,j}
a_{\mathbf q,j}\nonumber\\ &+&\sum_{\mathbf p \mathbf q,j\alpha\eta}
\frac{G^2}{2M\omega_{\mathbf q}}~c^\dagger_{\mathbf p,\alpha\eta}
c_{\mathbf p,\alpha\eta}\left(a^\dagger_{\mathbf q,j}a^\dagger_{-\mathbf q,j}
+a_{\mathbf q,j}a_{-\mathbf q,j}+2a^\dagger_{\mathbf q,j}a_{\mathbf q,j}
\right)\nonumber\\
&=&\sum_{\mathbf q,j}\omega_{\mathbf q} a^\dagger_{\mathbf q,j}a_{\mathbf q,j}
+\sum_{\mathbf q,j}\frac{G^2\rho}{2M\omega_{\mathbf q}}\left(a^\dagger_{\mathbf q,j}a^\dagger_{-\mathbf q,j}+a_{\mathbf q,j}a_{-\mathbf q,j}+2a^\dagger_{\mathbf q,j}a_{\mathbf q,j}\right)\nonumber\\
&=&\sum_{\mathbf q,j}\left(\omega_{\mathbf q}+\frac{G^2\rho}
{M\omega_{\mathbf q}}\right) a^\dagger_{\mathbf q,j}a_{\mathbf q,j}
+\sum_{\mathbf q,j}\frac{G^2\rho}{2M\omega_{\mathbf q}}
\left(a^\dagger_{\mathbf q,j}a^\dagger_{-\mathbf q,j}+a_{\mathbf q,j}
a_{-\mathbf q,j}\right)\nonumber\\
&=&\sum_{\mathbf q,j}\omega'_{\mathbf q} a^\dagger_{\mathbf q,j}a_{\mathbf q,j}
+\sum_{\mathbf q,j}\frac{g'}{2}\left(a^\dagger_{\mathbf q,j}
a^\dagger_{-\mathbf q,j}+a_{\mathbf q,j}a_{-\mathbf q,j}\right)
\end{eqnarray}
where $\omega'_{\mathbf q}=\left(\omega_{\mathbf q}+\frac{G^2\rho}{M\omega_q}
\right)=\omega_{\mathbf q}+g'$ with $g'=\frac{G^2\rho}{M\omega_{\mathbf q}}$. 

Now the equation of motion for the pions becomes
\begin{equation}
[H_{2\pi},\tilde a^\dagger_{\mathbf q,j}]=
\tilde\omega_{\mathbf q}\tilde a^\dagger_{\mathbf q,j}.
\end{equation}
This gives
\begin{equation}
\omega'_{\mathbf q}x_{\mathbf q}~a^\dagger_{\mathbf q,j}+g'x_{\mathbf q}~a_{\mathbf q,j}
-\omega'_{\mathbf q}y_q~a_{\mathbf q,j}-g'y_{\mathbf q}~a^\dagger_{\mathbf q,j}
=\tilde\omega_q(x_{\mathbf q}a_{\mathbf q j}
+y_{\mathbf q}a_{-\mathbf q j}^\dagger).
\end{equation}
The characteristic equation is
\begin{eqnarray}
\left|\begin{array}{cc}
(\omega'_{\mathbf q}-\tilde\omega_q)&-g'\\
g'&-(\omega'_{\mathbf q}+\tilde\omega_q)
\end{array}
\right|=\tilde\omega^2_{\mathbf q}-{\omega'_{\mathbf q}}^2+{g'}^2=0,
\end{eqnarray}
which gives
\begin{equation}
\tilde\omega_{\mathbf q}=\sqrt{{\omega'_{\mathbf q}}^2-{{g'}}^2}
\quad\quad
\mbox{with}\quad\quad
x_q=\sqrt{\frac{\omega'_q+\tilde\omega_q}{2\tilde\omega_q}}\quad\quad
y_q=\sqrt{\frac{\omega'_q-\tilde\omega_q}{2\tilde\omega_q}}\ .
\end{equation}
Now
\begin{equation}
H_{2\pi}=\sum_{\mathbf q,j}\tilde\omega_{\mathbf q} \tilde a^\dagger_{\mathbf q,j}\tilde a_{\mathbf q,j}+\frac{3}{2}\sum_{\mathbf q}(\tilde\omega_{\mathbf q}-\omega'_{\mathbf q})\ .
\end{equation}
We now have to include a term which corresponds to a
phenomenological repulsion energy between the pions of a ``pair'' in the 
above Hamiltonian $H_{2\pi}$ and is given by
\begin{equation}
H_m^R=A\sum_{\mathbf q,j}~e^{R_\pi^2\mathbf q^2}a^\dagger_{\mathbf q,j}a_{\mathbf q,j}
\end{equation}
where the two parameters $A$ and $R_\pi$ correspond to the strength and length
scale, respectively, of the repulsion and will be determined phenomenologically.
This term amounts to imposing a cut off on the momentum $\mathbf {q}$ which
accounts to the fact that momenta larger than $k_f$ are not dynamically
meaningful. With this repulsion term, we now have
\begin{equation}
\omega'_{\mathbf q}=\left(\omega_{\mathbf q}+A~e^{R_\pi^2\mathbf q^2}+
\frac{G^2\rho}{M\omega_{\mathbf q}}\right)=\omega_{\mathbf q}+A~e^{R_\pi^2
\mathbf q^2}+g'
\end{equation}
with
\begin{equation}
g'=\frac{G^2\rho}{M\omega_{\mathbf q}},\quad
\omega_{\mathbf q}=\sqrt{\mathbf q^2+m^2}\quad \mbox{and}\quad
\tilde\omega_{\mathbf q}=\sqrt{{\omega'_{\mathbf q}}^2-{{g'}}^2}\ .
\end{equation}
After transformation the Hamiltonian in equation (\ref{eqHam}) becomes
\begin{eqnarray}
\tilde H&\simeq&\sum_{\mathbf p,\alpha\eta}\varepsilon_\mathbf p~ 
c^\dagger_{\mathbf p,\alpha\eta}c_{\mathbf p,\alpha\eta}
+\sum_{\mathbf q,j}\tilde\omega_{\mathbf q} \tilde a^\dagger_{\mathbf q,j}
\tilde a_{\mathbf q,j}+\frac{3}{2}\sum_q(\tilde\omega_{\mathbf q}
-\omega'_{\mathbf q})\nonumber\\
&-&\sum_{\mathbf p\mathbf q,j\alpha\alpha'\eta\eta'}\frac{g_{\mathbf q}}
{\sqrt{\omega_{\mathbf q}V}}~c^\dagger_{\mathbf p+\mathbf q,\alpha\eta}
c_{\mathbf p,\alpha'\eta'}(i\mathbf\sigma\cdot\mathbf q)_{\alpha\alpha'}
\tau_j(\tilde a_{\mathbf q,j}+\tilde a^\dagger_{-\mathbf q,j})
\end{eqnarray}
where
\begin{equation}
g_{\mathbf q}= \frac{G~(x_{\mathbf q}-y_{\mathbf q})}{2M}\ .
\end{equation}
In the next section, we will consider RPA and calculate the correlation energy associated with one pion exchange for nuclear matter and neutron matter.
\section{Correlation energy associated with one pion exchange}
\subsection{nuclear matter}
We consider a Slater determinant of plane waves
\begin{equation}
|\Phi\rangle=\prod_{\alpha\eta,|\mathbf p|\leq p_F}
c^\dagger_{\mathbf p,\alpha\eta}|0\rangle,
\end{equation}
with $|0\rangle$ is the absolute vacuum, $p_F$ is the Fermi momentum and
\begin{equation}
c_{\mathbf p,\alpha,\eta}|0\rangle=0\ .
\end{equation}
Excitations with momentum transfer $\mathbf q$ are coupled to excitations with
momentum transfer $-\mathbf q$. Thus, the wave function $|\Psi\rangle$ which
describes such excitations should read
\begin{equation}
|\Psi\rangle=\exp S|\Phi\rangle\ ,
\end{equation}
where
\begin{eqnarray}
S_{\mathbf q j} &=&U_{\mathbf q}\sqrt{{\cal N}_q}\sum_{\alpha,\alpha',\eta,\eta'\mathbf p~\in~ \Omega_{\mathbf q}}
\langle \alpha, \eta|(\sigma\cdot\mathbf q) \tau_j|\alpha',\eta'\rangle ~
c^\dagger_{\mathbf p+\mathbf q,\alpha,\eta}c_{\mathbf p, \alpha',\eta'}\nonumber\\
&+&U_{-\mathbf q}\sqrt{{\cal N}_q}\sum_{\alpha,\alpha',\eta,\eta'\mathbf p~\in ~\Omega_{-\mathbf q}}
\langle \alpha,\eta|-(\mathbf\sigma\cdot\mathbf q)\tau_j|\alpha',\eta'
\rangle ~c^\dagger_{\mathbf p-\mathbf q,\alpha,\eta}c_{\mathbf p,\alpha',\eta'}\nonumber\\
&=&U_{\mathbf q}B_{\mathbf q, j}^\dagger+U_{-\mathbf q}B_{-\mathbf q, j}^\dagger\ .
\end{eqnarray}
In the above, ${\cal N}_q$ is the normalization factor insuring
\begin{equation}
{\cal N}_q\sum_{\alpha\alpha'\eta\eta',\mathbf p~\in~ \Omega_{\mathbf q}}|\langle
\alpha,\eta|\vec\sigma\cdot\mathbf q \tau_j|\alpha',\eta'\rangle|^2=4{\cal N}_q
\sum_{\mathbf p~\in~ \Omega_{\mathbf q}}\mathbf q^2=1
\end{equation}
and the domain $\Omega_q$ is defined by $|\mathbf p+\mathbf q| > p_F,~
|\mathbf p|\leq |p_F$. Only positive energy states are occupied. Here, 
$|\alpha,\eta\rangle$ denotes the spin iso-spin eigenstate, $\sigma_3|
\alpha,\eta\rangle=\alpha|\alpha,\eta\rangle$ and $\tau_3|\alpha,\eta
\rangle=\eta|\alpha,\eta\rangle$. The determination of the ${\cal N}_q$ 
is now very simple. All we need is the volume of the intersection of 2 
spheres of radius $p_F$, theirs centers being a distance $q$ apart. 
With this assumptions, we have
\begin{equation}
{\cal N}_q^{-1}=\frac{V}{(2\pi)^3}4\pi q^3\left(p^2_F-\frac{q^2}{12}\right)\ .
\end{equation}
The transformed unperturbed Hamiltonian becomes
\begin{equation}
H_0=\sum_{\mathbf p,\alpha\eta}\varepsilon_p~ c^\dagger_{\mathbf p,\alpha\eta}c_{\mathbf p,\alpha\eta}
+\sum_{\mathbf q,j}\tilde\omega_{\mathbf q} \tilde a^\dagger_{\mathbf q,j}
\tilde a_{\mathbf q,j}+\Delta\ ,\quad \mbox{where}\quad\Delta=
\frac{3}{2}\sum_q(\tilde\omega_q-\omega'_{\mathbf q})\ .
\end{equation}
The pion nucleon coupling reads,
\begin{equation}
H_{int}=-\sum_{\mathbf p\mathbf q,j\alpha\alpha'\eta\eta'}\langle\alpha\eta
|i(\vec\sigma\cdot\mathbf q)\tau_j|\alpha'\eta'\rangle\frac{g_{\mathbf q}}
{\sqrt{\omega_{\mathbf q}V}}~c^\dagger_{\mathbf p+\mathbf q,\alpha\eta}
c_{\mathbf p,\alpha'\eta'}(\tilde a_{\mathbf q,j}+
\tilde a^\dagger_{-\mathbf q,j})\ .
\end{equation}
In order to proceed, it is convenient to bosonize the Hamiltonian $H$, 
restricted to the subspace $S$. This is done by the replacement 
$S_{q,j}\rightarrow B_{q,j}$ satisfy boson commutation relations. 
The bosonized Hamiltonian reads as
\begin{equation}
H_B=E_{0}+\sum_{\mathbf q,j}\left(\varepsilon_q B^\dagger_{\mathbf q,j}
B_{\mathbf q,j} -Q_q(B^\dagger_{-\mathbf q,j}+
B_{\mathbf q,j})(\tilde a_{\mathbf q,j}+
\tilde a^\dagger_{-\mathbf q,j})+\tilde\omega_q 
\tilde a^\dagger_{\mathbf q,j}\tilde a_{\mathbf q,j}\right)\ .
\end{equation}
The parameters of $H_B$, namely $\varepsilon_q$ and $Q_q$ are fixed by 
expectation values such as
\[
\langle\Phi|B_{\mathbf qj}HB^\dagger_{\mathbf qj}|\Phi\rangle=
E_{HF}+\varepsilon_q,\quad\quad
\langle\Phi|\tilde a_{\mathbf qj}HB^\dagger_{\mathbf qj}|\Phi\rangle= Q_q,
\]
and
\begin{equation}
\langle\Phi|\tilde a_{\mathbf qj}B_{-\mathbf qj}H|\Phi\rangle=Q_q.
\end{equation}
The above Hamiltonian is diagonalized by a Bogoliubov transformation of the type
\begin{equation}
\Theta_{\mathbf qj}^{\dagger(n)}=x_1^{(n)}B^\dagger_{\mathbf qj}+x_2^{(n)}\tilde a^\dagger_{\mathbf qj}
+y_1^{(n)}B_{-\mathbf qj}+y_2^{(n)}\tilde a_{-qj},\quad\quad n=1,2
\end{equation}
which leads to excitation energies and the correlation energy.

In the above
\begin{eqnarray}
\varepsilon_q&=&~4~{\cal N}_q~q^2\sum_{\vec p\in~\Omega_q}\frac{1}{2M}
((\vec p+\vec q)^2-p^2)
=\frac{4{\cal N}_qq^2}{2M}\frac{V}{(2\pi)^3}\frac{4\pi q^2 p_F^3}{3}\nonumber\\
&=&\frac{1}{2M}\frac{4qp_F^3}{3(p_F^2-q^2/12)}\ ,
\end{eqnarray}
and
\begin{equation}
Q_q=\sqrt{\frac{{\cal N}_q^{-1}}{2\omega_q V}}g_q\ .
\end{equation}
The eigen frequencies are
\begin{equation}
\Omega_q^{(\pm)}=\frac{1}{\sqrt{2}}\sqrt{\varepsilon_q^2+\tilde\omega_q^2
\pm\sqrt{(\varepsilon_q^2-\tilde\omega_q^2)^2
+16\varepsilon_q\tilde\omega_qQ_q^2}}\ .
\end{equation}
The correlation energy becomes
\begin{equation}
E_{corr}=\frac{3}{2}\sum_{\vec q}\left(\Omega_q^{(+)}+\Omega_q^{(-)}-
\varepsilon_q-\tilde\omega_q\right)\ .
\end{equation}
\subsection{neutron matter}
The correlated Fermion wave function may be written, $\tau=-1$ for neutron and $\tau=1$ for proton
\begin{equation}
|\Psi\rangle=\exp S|\Phi\rangle,\quad\quad |\Phi\rangle=\prod_{\alpha,|\mathbf p|\leq p_F}c^\dagger_{\mathbf p,\alpha,-1}|0\rangle\ ,
\end{equation}
where the correlation operator reads
\begin{equation}
S=U_{\mathbf q}B_{\mathbf q,0}^\dagger+U_{-\mathbf q}B_{-\mathbf q,0}^\dagger+V_{\mathbf q}B_{\mathbf q, +}^\dagger+V_{-\mathbf q}B_{-\mathbf q,+}^\dagger\ ,
\end{equation}
with the quasi boson operators
\begin{eqnarray}
B_{\mathbf q, 0}^\dagger&=&\sqrt{{\cal N}_q}\sum_{\alpha,\alpha',\mathbf p~\in~ \Omega_{\mathbf q}}\langle \alpha, -1|(\sigma\cdot\mathbf q) \tau_0|\alpha',-1\rangle ~
c^\dagger_{\mathbf p+\mathbf q,\alpha,-1}c_{\mathbf p, \alpha',-1}\\
B_{\mathbf q, +}^\dagger&=&\sqrt{{\cal N}'_q}\sum_{\alpha,\alpha',|\mathbf p|~\leq~ p_F}\langle \alpha, 1|(\sigma\cdot\mathbf q) \tau_+|\alpha',-1\rangle ~
c^\dagger_{\mathbf p+\mathbf q,\alpha,1}c_{\mathbf p, \alpha',-1}\ ,
\end{eqnarray}
where $\tau_0=\tau_3$ and $\tau_+=(\tau_1+i\tau_2)/2$. Here, ${\cal N}_q$ and ${\cal N'}_q$ are normalization factors insuring
\begin{eqnarray}
{\cal N}_q\sum_{\alpha,\alpha',\mathbf p~\in~ \Omega_{-\mathbf q}}|\langle
\alpha,-1|\vec\sigma\cdot\mathbf q \tau_j|\alpha',-1\rangle|^2=2{\cal N}_q
\sum_{\mathbf p~\in~ \Omega_{\mathbf q}}\mathbf q^2&=&1\ ,\\
{\cal N'}_q\sum_{\alpha,\alpha',|\mathbf p|~\leq p_F}|\langle
\alpha,1|\vec\sigma\cdot\mathbf q \tau_j|\alpha',-1\rangle|^2=2{\cal N'}_q
\sum_{\mathbf p~\leq p_F}\mathbf q^2&=&1\ .
\end{eqnarray}
As earlier the domain $\Omega_q$ is defined by 
$|\mathbf p+\mathbf q| > p_F,~|\mathbf p|\leq |p_F$. 
The determination of the normalization, ${\cal N}_q$ and ${\cal N'}_q$, 
is now very simple. To compute ${\cal N}_q$, all we need is the volume 
of the intersection of two spheres of radius $p_F$, theirs centers 
being a distance $q$ apart. We found
\begin{equation}
{\cal N}_q^{-1}=\frac{V}{(2\pi)^3}2\pi q^3\left(p^2_F-\frac{q^2}{12}\right),
\quad\quad {\cal N'}_q^{-1}=\frac{V}{(2\pi)^3}\frac{8\pi}{3} q^2p^3_F\ .
\end{equation}
The kinetic energy for the particle-hole pairs with momentum $\mathbf q$ become
\begin{eqnarray}
\varepsilon_q&=&2~{\cal N}_q~q^2\sum_{\vec p\in~\Omega_q}
\frac{1}{2M}((\vec p+\vec q)^2-p^2)
=\frac{2{\cal N}_qq^2}{2M}\frac{V}{(2\pi)^3}\frac{4\pi q^2 p_F^3}{3}\nonumber\\
&=&\frac{1}{2M}\frac{2qp_F^3}{3(p_F^2-q^2/12)}\ ,\\
\varepsilon'_q&=&2~{\cal N}_q~q^2\sum_{\vec p\leq p_F}\frac{1}{2M}
((\vec p+\vec q)^2-p^2) =\frac{q^2}{2M}\ .
\end{eqnarray}
The pion nucleon interaction becomes
\begin{eqnarray}
H_{int}=&-&\sum_{\mathbf p\mathbf q,\alpha\alpha'\eta\eta'}
\langle\alpha\eta|i(\vec\sigma\cdot\mathbf q)\tau_0|\alpha'
\eta'\rangle\frac{g_{\mathbf q}}{\sqrt{\omega_{\mathbf q}V}}
c^\dagger_{\mathbf p+\mathbf q,\alpha\eta}c_{\mathbf p,\alpha'\eta'}
(\tilde a_{\mathbf q,0}+\tilde a^\dagger_{-\mathbf q,0})\nonumber\\
&-&\sum_{\mathbf p\mathbf q,\alpha\alpha'\eta\eta'}\langle\alpha
\eta|i(\vec\sigma\cdot\mathbf q)\tau_+|\alpha'\eta'
\rangle\frac{g_{\mathbf q}}{\sqrt{\omega_{\mathbf q}V}}
c^\dagger_{\mathbf p+\mathbf q,\alpha\eta}c_{\mathbf p,\alpha'\eta'}
(\tilde a_{\mathbf q,+}+\tilde a^\dagger_{-\mathbf q,-})\nonumber\\
&-&\sum_{\mathbf p\mathbf q,\alpha\alpha'\eta\eta'}
\langle\alpha\eta|i(\vec\sigma\cdot\mathbf q)\tau_-|\alpha'\eta'
\rangle\frac{g_{\mathbf q}}{\sqrt{\omega_{\mathbf q}V}}
c^\dagger_{\mathbf p+\mathbf q,\alpha\eta}c_{\mathbf p,\alpha'\eta'}
(\tilde a_{\mathbf q,+}+\tilde a^\dagger_{-\mathbf q,-})\ .\nonumber\\
\end{eqnarray}
The effective bosonized Hamiltonian containing two pion exchange becomes
\begin{eqnarray}
H_B=E_{0}&+&\sum_{\mathbf q}\left(\varepsilon_q B^\dagger_{q,0}B_{\mathbf q,0}
+\varepsilon'_q B^\dagger_{q,+}B_{\mathbf q,+}+
\tilde\omega_{\mathbf q} \sum_j\tilde a^\dagger_{\mathbf q,j}
\tilde a_{\mathbf q,j}\right)\nonumber\\
&-&\sum_{\mathbf q}Q_q(B^\dagger_{\mathbf q,0}+B_{\mathbf q,0})
(\tilde a_{\mathbf q,0}+\tilde a^\dagger_{-\mathbf q,0})\nonumber\\
&-&\sum_{\mathbf q}Q'_q\left(B^\dagger_{\mathbf q,+}(\tilde a_{\mathbf q,+}+\tilde a^\dagger_{-\mathbf q,-})+B_{\mathbf q,+}(\tilde a_{-\mathbf q,-}+\tilde a^\dagger_{\mathbf q,+})\right)\ ,
\end{eqnarray}
where
\begin{equation}
Q_q=\sqrt{\frac{{\cal N}_q^{-1}}{2\omega_q V}}~g_q,\quad\quad Q'_q=\sqrt{\frac{{\cal N'}_q^{-1}}{2\omega_q V}}~g_q\ .
\end{equation}
The RPA equations are easily obtained. For the modes with charge,
\begin{eqnarray}
&&[H_B,(X_qB^\dagger_{\mathbf q,+}+\zeta_q\tilde a^\dagger_{\mathbf q,+}+
\eta_q\tilde a_{-\mathbf q,-})]\nonumber\\
&=& B^\dagger_{\mathbf q,+}(X_q\varepsilon'_q+\zeta Q'_q-\eta_qQ'_q
+\tilde a^\dagger_{\mathbf q,+}(X_qQ'_q+\zeta_q\tilde\omega_q)+
\tilde a_{-\mathbf q,-}(X_qQ'_q-\eta_q\tilde\omega_q)\nonumber\\
&=&\Omega_q(X_qB^\dagger_{\mathbf q,+}+x_q\tilde a^\dagger_{\mathbf q,+}+
y_q\tilde a_{-\mathbf q,-}).
\end{eqnarray}
The characteristic is a cubic and becomes
\begin{equation}
\left|\begin{array}{ccc}
(\varepsilon'_q-\Omega_{\mathbf q})& Q'_q&-Q'_q\\
Q'_q&\tilde\omega_{\mathbf q}-\Omega_q&0\\
Q'_q&0&-\tilde\omega_{\mathbf q}-\Omega_q\end{array}
\right|=-\Omega^3_q+\varepsilon'_q\Omega_{\mathbf q}^2+
\tilde\omega_{\mathbf q}^2\Omega_{\mathbf q}+2{Q'}^2_q
\tilde\omega_{\mathbf q}-\varepsilon'_q\tilde\omega_{\mathbf q}^2=0
\end{equation}
where the solutions are

\begin{eqnarray}
\Omega^{(1)}_q &=& \frac{\varepsilon' }{3} 
+\frac{2^{1/3} \left(-{\varepsilon'}^2-3 \tilde\omega ^2\right)}
{3\left(-2 {\varepsilon'}^3-54 {Q'}^2 \tilde\omega +18 \varepsilon'  
\tilde\omega ^2+
\sqrt{4 \left(-{\varepsilon'}^2-3 \tilde\omega ^2\right)^3+\left(-2 
{\varepsilon'}^3-54 {Q'}^2 \tilde\omega +18 \varepsilon' 
\tilde\omega ^2\right)^2}\right)^{1/3}}\nonumber \\
&-&\frac{1}{3\times 2^{1/3}}
\left(-2 {\varepsilon'}^3-54 {Q'}^2 \tilde\omega +18 \varepsilon'  \tilde\omega ^2+
\sqrt{4 \left(-{\varepsilon'}^2-3 \tilde\omega ^2\right)^3+
\left(-2 {\varepsilon'} ^3-54 {Q'}^2 \tilde\omega +
18 \varepsilon'\tilde\omega ^2\right)^2}\right)^{1/3}\nonumber\\
\end{eqnarray}

\begin{eqnarray}
\Omega^{(2)}_q &=&\frac{\varepsilon' }{3}
-\frac{(1+i\sqrt{3}) \left(-{\varepsilon'}^2-3 \tilde\omega ^2\right)}
{3\times 2^{2/3}\left(-2 {\varepsilon'}^3-54 {Q'}^2 \tilde\omega +18 
\varepsilon'\tilde\omega ^2+
\sqrt{4 \left(-{\varepsilon'}^2-3 \tilde\omega ^2\right)^3+
\left(-2 {\varepsilon'}^3-54 {Q'}^2 \tilde\omega +18 \varepsilon'  
\tilde\omega ^2\right)^2}\right)^{1/3}}\nonumber \\
&+&\frac{1}{6\times 2^{1/3}}(1-i\sqrt{3})
\left(-2 {\varepsilon'}^3-54 {Q'}^2 \tilde\omega 
+18 \varepsilon'  \tilde\omega ^2+
\sqrt{4 \left(-{\varepsilon'}^2-3 \tilde\omega ^2\right)^3+
\left(-2 {\varepsilon'} ^3-54 {Q'}^2 \tilde\omega +
18 \varepsilon'\tilde\omega ^2\right)^2}\right)^{1/3}\nonumber\\
\end{eqnarray}

\begin{eqnarray}
\Omega^{(3)}_q &=&\frac{\varepsilon' }{3}
-\frac{(1-i\sqrt{3}) \left(-{\varepsilon'}^2-3 \tilde\omega ^2\right)}
{3\times 2^{2/3}\left(-2 {\varepsilon'}^3-54 {Q'}^2 \tilde\omega +18 
\varepsilon'\tilde\omega ^2+
\sqrt{4 \left(-{\varepsilon'}^2-3 \tilde\omega ^2\right)^3+
\left(-2 {\varepsilon'}^3-54 {Q'}^2 \tilde\omega +18 \varepsilon'  
\tilde\omega ^2\right)^2}\right)^{1/3}}\nonumber \\
&+&\frac{1}{6\times 2^{1/3}}(1+i\sqrt{3})
\left(-2 {\varepsilon'}^3-54 {Q'}^2 \tilde\omega 
+18 \varepsilon'  \tilde\omega ^2+
\sqrt{4 \left(-{\varepsilon'}^2-3 \tilde\omega ^2\right)^3+
\left(-2 {\varepsilon'} ^3-54 {Q'}^2 \tilde\omega 
+18 \varepsilon'\tilde\omega ^2\right)^2}\right)^{1/3}\nonumber\\
\end{eqnarray}
Similarly
\begin{equation}
[H_B,(Y_qB_{-\mathbf q,+}+\tilde\zeta_q\tilde a^\dagger_{-\mathbf q,+}
+\tilde\eta_q\tilde a_{\mathbf q,-})]=
\Omega_q(Y_qB_{-\mathbf q,+}+\tilde\zeta_q
\tilde a^\dagger_{-\mathbf q,+}+\tilde\eta_q\tilde a_{\mathbf q,-})\ .
\end{equation}
This leads to a similar equation with $\Omega_q$ replaced by $-\Omega_q$, 
so that the eigenfrequencies occur in pairs, $\pm\Omega_q$. The correlation 
energy for the charged modes becomes
\begin{equation}
E'_{corr}=\frac{1}{2}\sum_q(|\Omega^{(1)}_q|+|\Omega^{(2)}_q|+|\Omega^{(3)}_q|
-\varepsilon'_q-2\tilde\omega_q)\ .
\end{equation}
The eigen frequencies of uncharged modes are
\begin{equation}
\Omega_q^{(\pm)}=\frac{1}{\sqrt{2}}\sqrt{\varepsilon_q^2+\tilde\omega_q^2
\pm\sqrt{(\varepsilon_q^2-\tilde\omega_q^2)^2+16\varepsilon_q
\tilde\omega_qQ_q^2}}\ .
\end{equation}
The correlation energy becomes
\begin{equation}
E_{corr}=\frac{1}{2}\sum_{\vec q}\left(\Omega_q^{(+)}+\Omega_q^{(-)}
-\varepsilon_q-\tilde\omega_q\right)\ .
\end{equation}
which are given in earlier section.

\section{Results and Discussion}

We first proceed to describe the binding energy for nuclear matter.
We obtain the free nucleon kinetic energy density
\begin{equation}
h_f  = \langle\Phi|Tr[\hat \rho _N {\cal H}_N(\mathbf x)]|\Phi\rangle
={\gamma k_f^3 \over 6\pi^2} \left (M+{3\over 10}{k_f^2\over M}\right ).
\label{e15}
\end{equation}
In the above equation, spin degeneracy factor $\gamma$~=~4 (2) for nuclear 
matter (neutron matter) and , $k_f$ 
represents the Fermi momenta of the nucleons. The Fermi momenta
$k_f$ and the nucleon densities are related by
$k_f = ({6\pi^2\rho / \gamma})^{1 \over 3}$.
\begin{table}[!ht]
\begin{center}
\caption{Parameters of the model obtained self consistently at saturation 
density.}
\begin{tabular}{ccc}\\ \hline
a      &   $R_\pi$     &  \lomega    \\
(MeV)  &    (fm)       &   (fm$^2$)   \\
\hline
14.58  &   1.45 &   3.07   \\
\hline
\end{tabular}
\label{table1}
\vspace {-1cm}
\end{center}
\end{table}
It is well known that the short range interaction plays a crucial role in
determining the saturation density which is mediated by the
iso-scalar vector $\omega$ mesons. Here we introduce the energy of repulsion by
the simple form \cite{mishra90,mishra92}
\begin{equation}
h_\omega=\lomega\rho^2,\label{e26}
\end{equation}
where the parameter \lomega ~is to be fixed using the saturation
properties of nuclear matter as described in Ref.\cite{sarangi08}.
Thus we finally write down the binding energy per nucleon $E_B$ of the
symmetric nuclear matter (SNM):
\begin{equation}
E_B = {E_0\over\rho} - M  \label{e29}
\end{equation}
where
\begin{equation}
E_0=h_f+\frac{3}{2}\sum_q\tilde\omega_q+h_\omega\ .
\end{equation}
In the above, $E_0$ is the energy density of nuclear matter or neutron matter
without one pion correlation. The
expression for $E_0$ contains the three model parameters $a$, $R_\pi$, and 
\lomega~ as introduced in the earlier section. These parameters are 
determined self-consistently through the saturation properties of nuclear 
matter at saturation density $\rho_0$ = 0.15 fm$^{-3}$ with and without
the correlations. While pressure $P$ vanishes at saturation
density for symmetric nuclear matter, the values of binding energy per nucleon
are chosen to be $-16$~MeV. In the numerical calculations, we have used the 
nucleon mass $M=940$ MeV, the pion masses
$m=140$ MeV and the omega meson mass, $m_\omega=783$ MeV, and the 
$\pi-N$ coupling constant $G^2/4\pi=14.6$. 

\begin{figure}[!ht]
\begin{center}
\includegraphics[width=3.4in,height=3.6in]{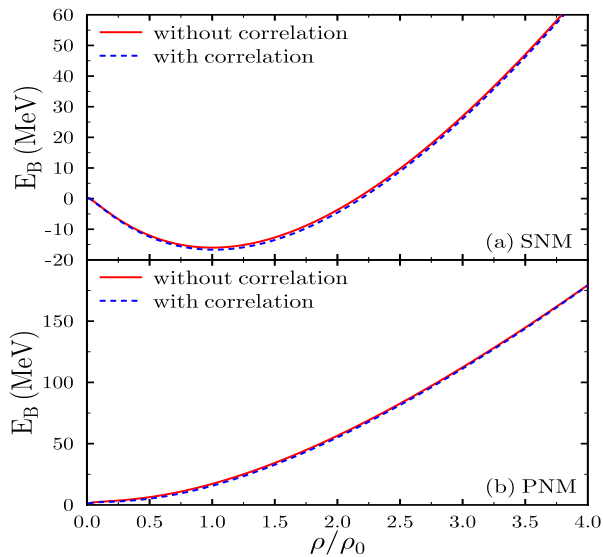}
\vspace {-1cm}
\caption{Binding energy of symmetric nuclear matter (SNM) and pure neutron 
matter (PNM). The correlation is related to one-pion exchange}
\label{fig:eb}
\end{center}
\end{figure}
\begin{figure}[!ht]
\begin{center}
\includegraphics[width=3.4in,height=3.6in]{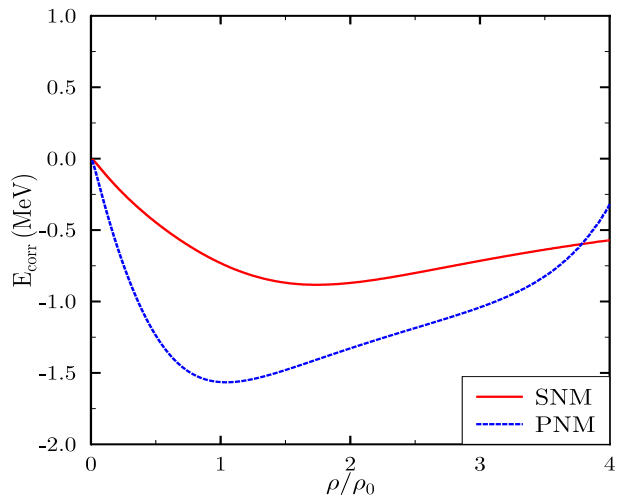}
\vspace {-2cm}
\caption{The correlation energy from one-pion exchange in symmetric 
nuclear matter (SNM) and in pure neutron matter (PNM)}
\label{fig:corr}
\end{center}
\end{figure}
We now discuss the results obtained in our calculation. We first construct a
Bogoliubov transformation for the pion pairs operators and calculate the
energy associated with it. We next calculate the correlation energy due to the
one pion exchange in nuclear matter and neutron matter at RPA using
the generator coordinate method. The binding 
energy per nucleon $E_B$ as a function of the density of the system is often 
refered as the nuclear equation of state (EOS). In figure 1, we present the 
EOS with and without correlation for the nuclear matter and for neutron matter.
The correlation is related to one pion exchange. As expected, the binding 
energy for nuclear matter with and without correlation initially decreases 
with density and reaches a minimum at $\rho/\rho_0=1$ and then increases.

\begin{figure}[!ht]
\begin{center}
\includegraphics[width=4.0in,height=3.6in]{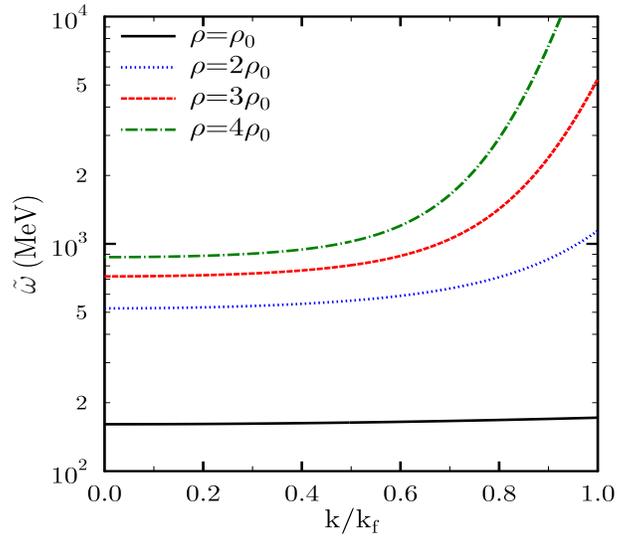}
\vspace {-2cm}
\caption{The dispersion relation with the quantum number of the pions, 
$\tilde\omega$ for nuclear matter}
\label{fig:omegatilde}
\end{center}
\end{figure}
\begin{figure}[!ht]
\begin{center}
\includegraphics[width=4.0in,height=3.6in]{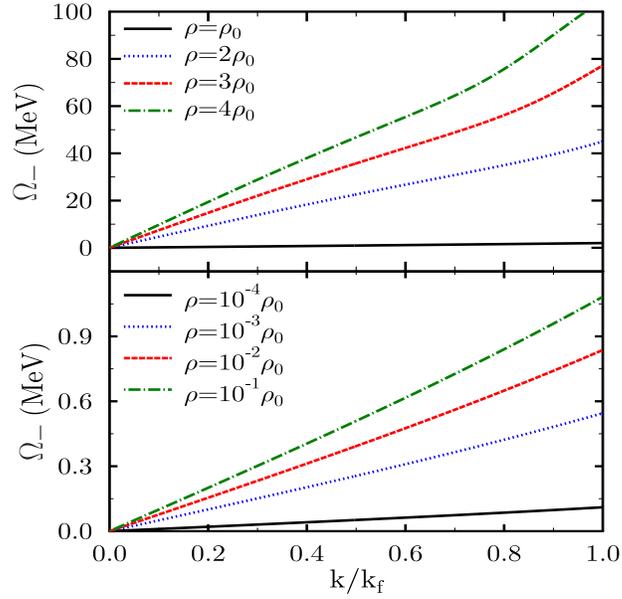}
\vspace {-1cm}
\caption{The dispersion relation of the RPA mode with the quantum number of 
the pions, $\Omega_-$, at high densities (upper panel) and for low densities
(lower panel) for nuclear matter. This corresponds to zero sound modes.}
\label{fig:omegamin}
\end{center}
\end{figure}
In figure 2, we show the variation of the correlation energy, $E_{corr}$ as
a function of density for symmentric nuclear matter (SNM) and for the pure
neutron matter (PNM). The correlation energy initially decreases with density
and then increses after the saturation density. The correlation energy for
the neutron matter gives larger as compared to the nuclear matter at different
densities.

Dispersion relation of modes with the quantum numbers of the pions 
in nuclear medium is an interesting aspect. In 
figure 3, we plot the dispersion relations arisig from the two pion
coherent states versus with the momentum at different densities. The increase 
of the pion dispersion relation for $k/k_f ~> ~0.6$ is probably an
artifact of the repulsion term of equation (26).
\begin{figure}[!ht]
\begin{center}
\includegraphics[width=3.8in,height=3.2in]{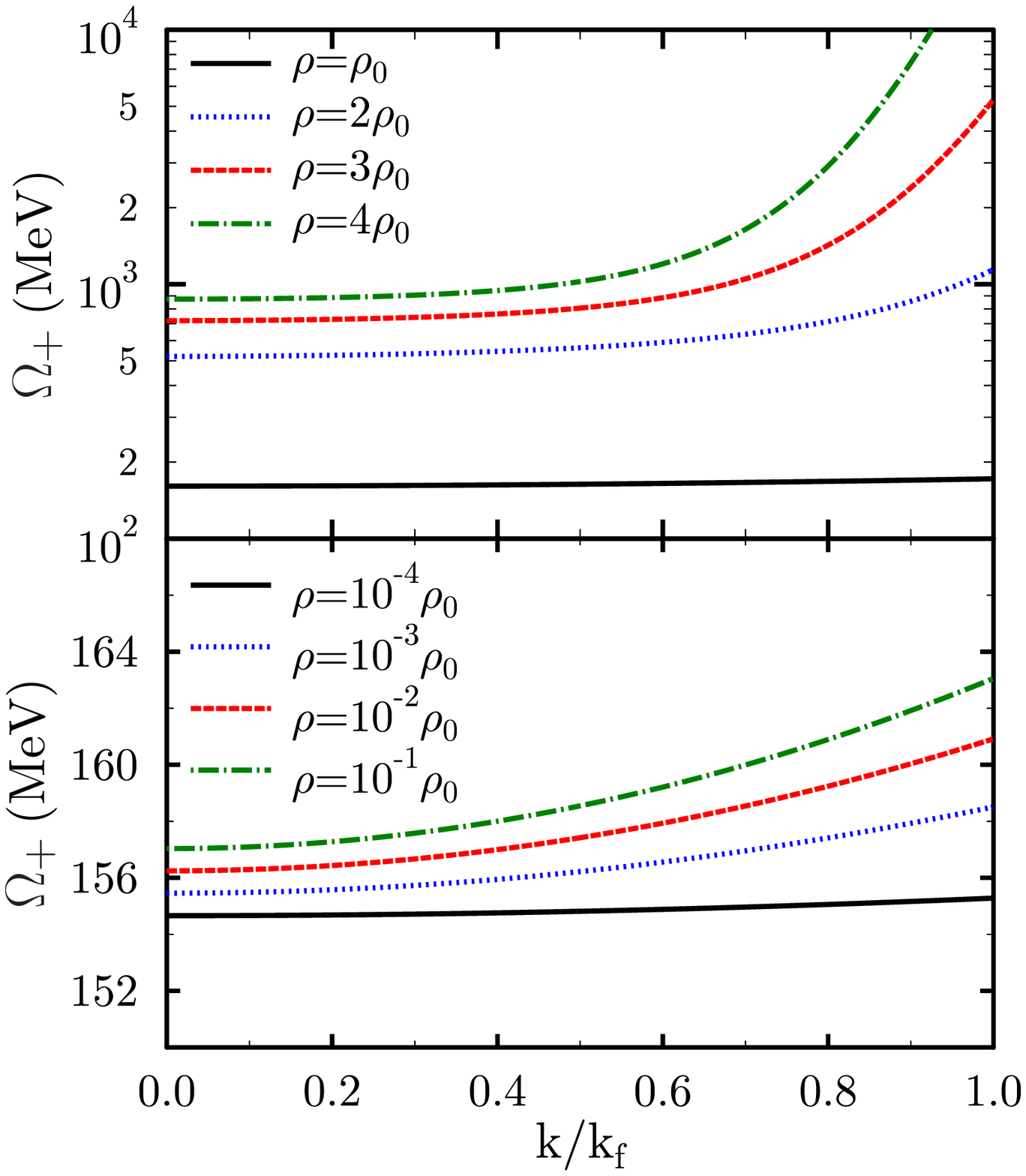}
\vspace {-1cm}
\caption{The dispersion relation of the RPA mode with the quantum number of 
the pions, $\Omega_+$, at high densities (upper panel) and at low densities
(liwer panel) for nuclear matter. It is showing an increase with density
of the effective mass of the pions.}
\label{fig:omegaplus}
\end{center}
\end{figure}
\begin{figure}[!ht]
\begin{center}
\includegraphics[width=3.8in,height=3.2in]{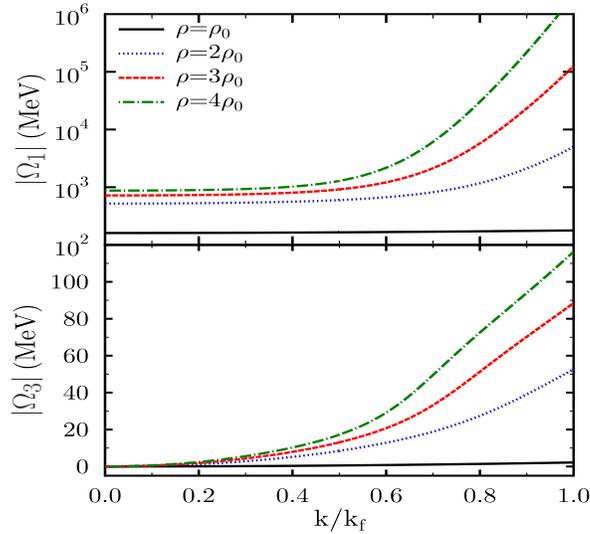}
\vspace {-1cm}
\caption{The dispersion relation of the RPA mode with the quantum number 
of the pions, $|\Omega_1|$ and $|\Omega_3|$, for neutron  matter}
\label{fig:omega2}
\end{center}
\end{figure}

For nuclear matter, we observed two RPA modes with quantum number of the pions,
$\Omega_\pm$. In figure 4, we have shown the dispersion relation for the
RPA modes, $\Omega_-$, versus momentum at different densities for nuclear 
matter. At $\rho=\rho_0$, $\Omega_-$ increases very slowly with momentum.
However at $\rho=4 \rho_0$, it increases fast. In the lower panel of the 
figure 4, we have plotted $\Omega_-$ for smaller densities. At small 
densities, $\Omega_-$ increases monotonically with momentum $k$ and
corresponds to zero sound modes.

In figure 5, we have shown the dispersion relation for the
RPA modes, $\Omega_+$, versus momentum at different densities for nuclear
matter. It is found the the magnitude of the $\Omega_+$ is larger
compared to $\Omega_-$. In lower panel of the figure 5, we plotted $\Omega_+$
for smaller densities, showing an increase with density of the
effective mass of the pions.

We next study the RPA modes for the neutron matter. There are three modes
for the charge pions.  In figure 6, we
plot the RPA frequencies versus momentum $k$ for neutron matter. The 
$|\Omega_1|$ and $|\Omega_2|$ are equal. In the upper pannel, we show 
$|\Omega_1|$ versus momentum $k$. All the three RPA frequencies increase 
with density. In the lower panel, we plot $|\Omega_3|$ versus momentum 
$k$ which corresponds to zero sound modes. It is found that there is no
sign of pion condensation in higher densities with the RPA modes. 

In conclusion, we have derived a pion nucleon Hamiltonian in a non-relativistic
limit. We then have constructed a Bogoliubov transformations for the pion pair
operators and calculate the energy associated with the pion pairs for the
nuclear matter and neutron matter. This is an extension of the mean field 
approach of Walecka where the classical fields are replaced by the quantum 
coherent states for the pion pairs. We then calculated the correlation
energies due to one pion exchange in nuclear matter and neutron matter
at RPA using generator coordinate method. It is found that there is no 
sign of pion condensation in higher densities with the RPA modes.\\

\noindent{\bf Acknowledgements}\\

One of the author (PKP) thanks the hospitality and the friendly atmosphere
provided to him during the stay at Departmento de Fisica, Universidade Coimbra.
This work was partially supported by FEDER and FCT (Portugal) under the
projects PDCT/FP/64707/2006 and CERN/FP/83505/2008.\\

\end{document}